\documentclass[a4paper]{article}

\usepackage{INTERSPEECH2020}
\usepackage{amsmath,graphicx}
\usepackage{multirow}

\usepackage{float}

\usepackage{bm}
\newcommand{\VEC}[1]{\ensuremath{\bm{#1}}}
\newcommand{\vecX}{\VEC{X}}
\newcommand{\vecY}{\VEC{Y}}

\newcommand{\WavAE}{\ensuremath{\text{Wav}_{\text{AE}}}~}

\newcommand{\WavCCCAE}{\ensuremath{\text{Wav}_{\text{CCCAE}}}~}

\newcommand{\ModelMFCC}{\ensuremath{\text{M}_{\text{{MFCC}}}}~}
\newcommand{\ModelAE}{\ensuremath{\text{M}_{\text{{AE}}}}~}

\newcommand{\ModelCCCAE}{\ensuremath{\text{M}_{\text{{CCCAE}}}}~}

\newcommand{\MyVal}[1]{--}

\newcommand{\etal}{\mbox{\emph{et al.\ }}}

\title{Prediction Of Head Motion From Speech Waveforms With A Canonical-Correlation-Constrained Autoencoder}
\name{JinHong Lu, Hiroshi Shimodaira}
\address{Centre for Speech Technology Research, School of Informatics,\\ University of Edinburgh, United Kingdom}
\email{jinhong.l@sms.ed.ac.uk, H.Shimodaira@ed.ac.uk}

\begin{document}

\maketitle
\begin{abstract}
 This study investigates the direct use of speech waveforms to
 predict head motion for speech-driven head-motion synthesis, whereas the use of spectral features such as MFCC as basic input features together with additional features such as energy and F0 is common in the literature. We show that, rather than combining
 different features that originate from waveforms, it is more
 effective to use waveforms directly predicting corresponding head
 motion. The challenge with 
 the waveform-based approach is that waveforms contain a large
 amount of information irrelevant to predict head motion,
 which hinders the training of neural networks. To overcome the
 problem, we propose a canonical-correlation-constrained autoencoder (CCCAE), 
 where hidden layers are trained to not only minimise
 the error but also maximise the canonical correlation with head
 motion. 
 Compared with an MFCC-based system, the proposed system shows
 comparable performance in objective evaluation, and better
 performance in subject evaluation.
\end{abstract}
\noindent\textbf{Index Terms}: head motion synthesis, speech-driven animation, deep canonically correlated autoencoder

\section{Introduction} \label{sec:intro}
Head motion such as nodding and shaking is an important nonverbal
communication channel in human-human communication. In addition to
the head motion as nonverbal signals, Hadar \etal
\cite{hader-steiner:language-speech:1983} have shown another type of head motion that is directly related to speech production.
It is essential
for animated talking heads to realise both types of natural head motion as well as
lip-sync to make the avatar more 
human-like. Compared with lip-sync, the synthesis of head motion from
audio speech is more challenging, since the link between speech
and head motion is less clear, and not only speech, but also various
factors such as emotion, intention, and stance are involved.

The present study considers a link between head motion and
acoustic speech features, whose original representation is given as
acoustic waveform signals, and seeks compact and
efficient representation of speech features to predict corresponding
head motion.
Kuratate \etal \cite{Kuratate1999} found that fundamental frequency (F0)
and head motion had a correlation of 0.83 at sentence-level.
Busso \etal \cite{Busso2007} also confirmed a strong sentence-level
correlation (r=0.8) between head motion and mel-frequency cepstral
coefficients (MFCCs), where data was recorded for an actor reading the
scripts of short sentences. 

As we show in experiments, it is a different scenario in
natural conversations, where there is a much larger degree of
variation in head motion and we cannot find such strong correlations.
A similar observation is reported for a dialogue corpus by Sadoughi \etal
\cite{Najmeh2017}, where they have found a global canonical correlation analysis (global CCA) of 0.1931 between the original head
movements and speech (F0 and energy).

In order to tackle the problem of a weak link between speech and head
motion, other features and their combination have been explored.
Ben-Youssef \etal\cite{BenYoussef2014} found that the articulatory
features or EMA features that were estimated from speech were more
useful to predict head motion.
Ding \etal\cite{Ding2015} examined LPC,
MFCC, and filter bank (FBank) features and showed 
that FBank-based system outperformed MFCC-based one.
Haag \etal\cite{Haag2016} combined MFCC and EMA features to build bottleneck
features, which were then fed to DNN-BLSTM to predict head motion.
Greenwood \etal\cite{Greenwood2017} proposed CVAE-BLSTM and used the decoder as a generative model to predict head motion, where the FBank features were used as the condition.

The purpose of using a combination of different features in the
previous studies was to use richer information (e.g., prosodic
features) to train models and predict head motion.
Since all the acoustic features described above are derived from
raw speech waveforms, it is natural to consider the original
waveforms as the input to neural networks, so that we will be able to
fully make use of the information in the original observations.
So far, no one has investigated the use of original raw
waveforms to predict head motion. This is mainly because of (1) the high
dimensionality of raw waveform signals, which slows down the training of neural networks and requires high capacity in the
hardware support; (2) a large amount of irrelevant information
to predict head motion, which hinders the training of neural networks.

To overcome the problems of high dimensionality and irrelevant
information, we propose a canonical-correlation-constrained autoencoder (CCCAE) to extract
low-dimensional features 
from raw waveforms, where hidden layers are trained not only to minimise
the error of encoding and decoding, but also
maximise the canonical correlation with head motion. The extracted
features of a low dimension are then fed to another neural network for
regression to predict head motion.
We show that the features obtained with the proposed approach are more
useful for head-motion prediction than those with a standard
autoencoder.
We evaluate the new approach through comparisons with other acoustic
features in terms of objective and subjective measures.

\begin{figure}[h!tb]
  \centering
  \includegraphics[width=\columnwidth, height=8cm, keepaspectratio]{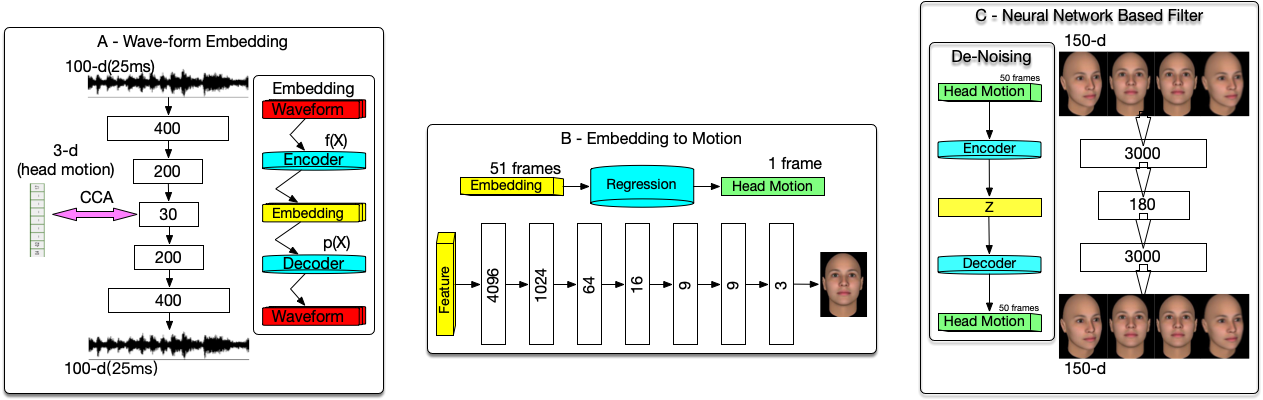}
  \vskip -3mm
  \caption{%
    Overview of the proposed system comprised of three modules:
    (A) waveform embedding with CCCAE, (B) DNN-based head motion regression
    from the embedded features, (C) post filter with an autoencoder.
  }
  
  \label{fig:proposed_model}
   \vskip -3mm
\end{figure}

\smallskip

\subsection{Relation to prior work}
Using raw waveforms for acoustic modelling with neural networks is one
of the active areas in automatic speech recognition %
\cite{Sainath2015,Ghahremani2016,Tske2018,platen2019}.
However, to the best of our knowledge, no one has investigated the use of raw
waveforms for speech-driven head-motion synthesis, in which a
set of two data streams, speech and head motion, is dealt with rather
than a single stream of speech.
Chandar \etal\cite{ChandarKLR15} and Wang \etal\cite{WangALB16} have proposed the
framework of correlational neural networks and deep canonically
correlated autoencoder (DCCAE), respectively, to effectively model two data streams, and they applied the models to 
cross-language tasks and multi-view feature learning, where
you can expect reasonably high correlations between two data streams.
The present study is different in
that the correlation between speech and head-motion features are much
weaker, and our proposed model employs only one autoencoder
whereas they employ two.

\section{Proposed System}

Our proposed system can be separated into three modules; (1) a
canonical-correlation-constrained autoencoder (CCCAE) for compressing the
high-dimensional waveform input to distributed embedding of low
dimensions; (2) a regression model for predicting the head motion from
the compressed embedding; (3) a post-filtering autoencoder for
reconstructing smooth head motion.
The overall framework of our proposed model is shown in Figure~\ref{fig:proposed_model}.

\subsection{Waveform Embedding}
The framework of autoencoder for a set of two data streams is
proposed by Chandar \etal\cite{ChandarKLR15} and Wang \etal\cite{WangALB16}.
DCCAE~\cite{WangALB16} consists of two autoencoders and optimises the
combination of canonical correlation between the learned "bottleneck"
representations and the reconstruction errors of the autoencoders.
Since head motion is parameterised with a time series of rotation 
vectors of three dimensions in the present study, we do not need to use an autoencoder to
reduce the dimensionality further. We thus employ a single autoencoder, in which hidden layers are trained in such a way as to not only minimise the reconstruction error, but also maximise the canonical correlation with head motion.
Thus, instead of projecting the two
features to a common subspace, we project raw waveforms to a subspace
so that the embedded features are well correlated with head motion.
We do not consider more advance architectures such as VAE/CVAE\cite{Greenwood2017, kingma2013autoencoding, sohn2015}, because standard AE is more effective in this task as the generative models, VAE/CVAE, is harder in the training. 

We train the proposed CCCAE with the following objective function,
\begin{align}
 \text{Obj}_{\text{CCCAE}} =  \sum_t \|\vecX_t - p(f(\vecX_t))\|^2 - \alpha\,\text{CCA}\left(f(\vecX), \vecY\right)
\end{align}
where $\vecX_t$ represents the input raw waveform vector at a time
instance $t$ to the encoder, $f(\,)$ represents the projection with the encoder,
$p(\,)$ represents the reconstruction with the decoder, $\vecX$ and $\vecY$
denote the whole sequences of waveform vectors and head motion
vectors, respectively, and
$\text{CCA}(\,)$ is the canonical correlation function.
%
$\alpha \ge 0 $ is the weighting factor, where $\alpha=0$ corresponds
to a standard autoencoder with an MSE loss function. 

\subsection{Head motion regression}
A simple feed-forward deep neural network is applied here for the
regression from the waveform embedded features to head motion. We do not
consider more complex models such as CNN and LSTM, because the present study focuses on a compact and efficient representation of speech features rather than the regression of head motion; and previous 
studies \cite{Ding2015, Haag2016} showed no large differences among the models.
Accordingly, we also do not consider auto-regressive models such as
WaveNet~\cite{Oord_Zen2016}.

As is shown in Figure~\ref{fig:proposed_model}(B), a context window of $\pm
25$ frames, which is equivalent to 525ms effective speech content, is 
employed to predict head motion parameters.

\subsection{Post-filter}
Since the output trajectories of the neural networks are noisy or
discontinuous due to the nature of speech, it is common to apply a post-filter as
post-processing to obtain smooth head motion trajectories for
animation~\cite{Haag2016, Ding2015BLSTMNN, Najmeh2018}.

The common training procedure of the de-noising model, which comprises applying dropout/Gaussian noise to the clean data for recreating noisy data~\cite{GhoshSAH17, Vincent:2010:SDA:1756006.1953039}, does not work with our model as 
the Gaussian noise method does not give the expected sudden and quick movements as they would naturally occur. 
The dropout method, on the other hand, drops one of the three trajectories of the head motion, 
and this strictly limits the movement, causing unnatural behaviour. 
Therefore, instead of removing the noise from the jerky head motion, 
we expect the de-noising filter to learn and know how the smooth head motion over a period should be. 
We assume a complete head motion in every consecutive 500ms~\cite{Hofer2007AutomaticHM} time frame, 
as the input, $M_f$, to the de-noising filter and the output are of the same length. 
We built a neural network based de-noising autoencoder
following the architecture, trained with the "clean"
data~\cite{jin2019}. 

\section{Experiments and Results}
\label{sec:experiment}
\subsection{Dataset}
We used the University of Edinburgh Speaker Personality and Mocap
Dataset~\cite{Haag2015TheUO}. This dataset contains expressive
dialogues of 13 native English semi-professional actors in
extroverted and introverted speaking styles and the dialogues are
non-scripted and spontaneous. For the purpose of our experiments, we
selected data from one male (Subject A) and one female (Subject
B). Six recordings (around 30 minutes) of each subject were
used for training, two (around 10 minutes) for
validation, and the remaining two (around 10 minutes) for evaluation,
ignoring the differences in terms of the speaking
style. Moreover, we have chosen the benchmark dataset, IEMOCAP~\cite{Busso2007}, to validate our model.
We selected the recordings of the first female subject from the dataset as Sadoughi\etal\cite{Najmeh2018}.
We trained our models for each subject. Note that
speaker-dependent training is a common practice in speech-driven head
motion synthesis~\cite{Busso2005,Ding2015,Najmeh2018}. 

\subsubsection{Speech Features} Audio in the database was recorded with a
headset microphone at 44.1 kHz with 32-bit depth and a MOTU 8pre
mixer~\cite{MOTU}. Separate recording channels were used for the two
speakers and a synchronisation signal was recorded on a third channel
in the mixer. For the purpose of this work, the audio signal was
downsampled to 4 kHz prior to feature extraction. Raw waveform vectors were
extracted using 25 ms windows
with 10 ms shifting, which resulted in 100 dimensions.
13 MFCCs feature is
formed by combining 1 energy coefficient and 12 Mel-cepstral coefficients,
using SPTK~\cite{SPTK}. 
We also added their first and second-order derivatives, resulting in 39-MFCCs.
Voicing probability and energy were computed using
openSMILE~\cite{Eyben2010}, and smoothed with a moving average filter
with a window length of 10 frames.
All the features were normalised in terms of variance for each dimension.

\subsubsection{Head Motion Features} Movements of the head as a 3D rigid-body
were recorded with the NaturalPoint Optitrack~\cite{NaturealPoint}
motion capture system at a 100 Hz sampling rate. From the marker coordinates, rotation matrices for the head motion were computed using singular value decomposition~\cite{Soderkvist1994}, which were further converted to rotation vectors of three dimensions.

\subsection{Experimental Setups}
We conducted preliminary experiments to decide the depth and width of
the models, which are shown in
Figure~\ref{fig:proposed_model}. 
We tested different numbers of nodes, 15, 30, and 60, for the embedding
layer of CCCAE, and decided to use 30 nodes based on the performance of the autoencoder.
In training and objective evaluation, we only used the frames where the target speaker for head-motion prediction was speaking, so that the models learnt the
relationship between speech and head motion properly.
In subjective evaluation, we made use of all the input audio sequences to generate
head motion parameters.
The following notations are used in the rest experiments.
\begin{itemize}
    \itemsep0em
      %
  \item \WavAE: Embedded features extracted from
    the standard autoencoder (i.e., the output of proposed CCCAE with $\alpha=0$)
  \item \WavCCCAE:  Embedded features extracted from
    the proposed CCCAE with $\alpha=1$
  \item \ModelMFCC: Regression model trained with MFCC feature
  \item \ModelAE: Regression model trained with \WavAE
  \item \ModelCCCAE: Regression model trained with \WavCCCAE
\end{itemize}
\ModelMFCC, \ModelAE, and \ModelCCCAE use the same
architecture in Figure~\ref{fig:proposed_model}(B) to predict head
motion, while each model takes different feature vectors as input.

Training was conducted on a GPU machine and a multi-CPU machine with
Tensorflow version 1.12 by mini-batch training using Adam optimisation
(learning rate 0.0002)~\cite{Diederik2014}.
We also employed layer-wide pre-training~\cite{Takaki2016}.


\begin{table}[tb]
\centering
\caption{%
  Local CCA between speech features and original head motion.
}
\begin{tabular}{|c|c|c|c|c|}
\hline
\multicolumn{1}{|c}{\multirow{2}{*}{Feature}}&\multicolumn{1}{|c|}{\multirow{2}{*}{Subject}}&\multicolumn{3}{c|}{CCA} \\
\cline{3-5}
\multicolumn{1}{|c}{}  &\multicolumn{1}{|c|}{}& Training & Valid & Test\\
\hline
\multicolumn{1}{|c|}{\multirow{2}{*}{F0+Energy}} & A&\MyVal{0.025}&\MyVal{0.111}&$0.107$      \\\cline{2-5} 
\multicolumn{1}{|c|}{}                      & B      &\MyVal{0.088}&\MyVal{0.107}& $0.117$  \\\cline{2-5} 
\multicolumn{1}{|c|}{}                      & IEMOCAP      &\MyVal{0.313}&\MyVal{0.342}& $0.301$  \\\hline
\multicolumn{1}{|c|}{\multirow{2}{*}{FBank}} & A      &\MyVal{0.155}&\MyVal{0.147}&   $0.143$   \\\cline{2-5} 
\multicolumn{1}{|c|}{}                      & B      &\MyVal{0.149}&\MyVal{0.167}& $0.157$     \\\cline{2-5} 
\multicolumn{1}{|c|}{}                      & IEMOCAP      &\MyVal{0.551}&\MyVal{0.462}& $0.392$  \\\hline
\multicolumn{1}{|c|}{\multirow{2}{*}{MFCC}} & A      &\MyVal{0.249}&\MyVal{0.224}&$0.238$      \\\cline{2-5} 
\multicolumn{1}{|c|}{}                      & B      &\MyVal{0.228}&\MyVal{0.247}&$0.257$      \\\cline{2-5} 
\multicolumn{1}{|c|}{}                      & IEMOCAP      &\MyVal{0.804}&\MyVal{0.783}& $0.784$  \\\hline
\multicolumn{1}{|c|}{\multirow{2}{*}{waveform}} & A      &\MyVal{0.210} &\MyVal{0.184}&$0.186$      \\\cline{2-5} 
\multicolumn{1}{|c|}{}                      & B      & \MyVal{0.097} & \MyVal{0.119} &    $0.157$  \\\cline{2-5} 
\multicolumn{1}{|c|}{}                      & IEMOCAP      &\MyVal{0.603}&\MyVal{0.596}& $0.658$  \\\hline
\multicolumn{1}{|c|}{\multirow{2}{*}{\WavAE}} & A  &$0.221$&$0.196$&  $0.196$   \\\cline{2-5} 
\multicolumn{1}{|c|}{}                      & B      &  $0.110$     &   $0.135$    & $0.176$ \\\cline{2-5} 
\multicolumn{1}{|c|}{}                      & IEMOCAP      &$0.626$&$0.626$& $0.689$  \\\hline
\multicolumn{1}{|c|}{\multirow{2}{*}{\WavCCCAE}} & A      &$0.264$&$0.234$&$0.248$      \\\cline{2-5} 
\multicolumn{1}{|c|}{}                      & B      &$0.220$&$0.240$&$0.266$      \\\cline{2-5}
\multicolumn{1}{|c|}{}                      & IEMOCAP      &$0.774$&$0.784$& $0.838$  \\
\hline
\end{tabular}
\vskip -2mm
\label{tab:feature_cca}
\end{table}

\begin{table}[tb]
\centering
\caption{%
  Comparison of different systems in terms of performance of head motion
  prediction, where NMSE and local CCA are calculated between
  predicted head motion and ground truth.
}

\begin{tabular}{|l|c|c|c|c|c|}
\hline
\multicolumn{1}{|c}{\multirow{2}{*}{System}}&\multicolumn{1}{|c|}{\multirow{2}{*}{Subject}}&\multicolumn{2}{c|}{Training}&\multicolumn{2}{c|}{Test} \\
\cline{3-6} 
\multicolumn{1}{|c}{}  &\multicolumn{1}{|c|}{}& NMSE & CCA & NMSE & CCA\\
\hline
\multicolumn{1}{|c|}{\multirow{2}{*}{\ModelMFCC}} & A      &  $0.82$&$0.47$ &  $1.58$&$0.33$\\\cline{2-6}
\multicolumn{1}{|c|}{}                      & B      &  $0.55$&$0.57$ &  $1.71$&$0.34$\\\cline{2-6} 
\multicolumn{1}{|c|}{}                      & IEMOCAP      & $0.46$ &$0.62$&$1.30$&$0.43$\\\hline
\multicolumn{1}{|c|}{\multirow{2}{*}{\ModelAE}} & A      &  $1.01$&$0.18$ & $1.07$&$0.20$\\\cline{2-6} 
\multicolumn{1}{|c|}{}                      & B      &   $1.21$    &   $0.09$    &    $1.16$  &$0.08$\\\cline{2-6} 
\multicolumn{1}{|c|}{}                      & IEMOCAP      & $0.98$ &$0.01$&$1.11$&$0.00$\\\hline
\multicolumn{1}{|c|}{\multirow{2}{*}{\ModelCCCAE}} & A      &  $0.58$&$0.42$ & $1.52$&$0.29$ \\\cline{2-6} 
\multicolumn{1}{|c|}{}                      & B      &  $0.69$&$0.39$ & $1.33$&$0.24$  \\\cline{2-6} 
\multicolumn{1}{|c|}{}                      & IEMOCAP      & $0.68$ &$0.61$&$1.14$&$0.37$\\
\hline
\end{tabular}
\vskip -5mm
\label{tab:rms_cca}
\end{table}

\smallskip
\subsection{Objective Evaluation}
To measure the similarity between two sequences of vectors,
we employed normalised mean-squared error (NMSE), where MSE is
normalised by the variance of ground truth, and local CCA~\cite{Haag2016}.
As opposed to global CCA, which calculates canonical correlations over
the whole sequence, local CCA calculates CCA scores for every
sub-sequence obtained with a time window and takes 
the average of the resulting scores. We used local CCA rather than global CCA, 
because head motion trajectories are not stationary and linear
correlations rarely hold over long periods. 
We used a time window of 300 frames or 3 seconds.

In addition to the speech features described before, for comparison purposes,
we also used F0+Energy (6 dimensions with delta and delta delta
features)~\cite{Najmeh2017},
FBank (27 dimensions of 26 filter-bank channels and log energy),
and waveform (100 dimensions), which is the input to the proposed CCCAE.

\smallskip
\subsubsection{CCA between speech features and original head motion}
Before evaluating the performance of head-motion prediction from speech, 
we carried out a basic correlation analysis between
speech features and 
head motion in terms of local CCA.
Table~\ref{tab:feature_cca} shows local CCA for each speech feature and for
each subject. Note that CCA scores on training and validation sets are
not shown for those features in which training is not involved.
It can be found that
F0+Energy gives the smallest, and MFCC and \WavCCCAE %
achieve the largest CCA scores with head motion.
Compared to waveform, we can see a large improvement on the test set
(by 33\% for Subject A and 69\% for Subject B) with \WavCCCAE, whereas
there is a small improvement with \WavAE. 
Looking at the results of the benchmark dataset, 
our proposed feature has a improvement of 15\% and 6.9\% to waveform and MFCC respectively.

%
\subsubsection{Evaluation of predicted head motion from speech}
Based on the result of the basic analysis, we chose the three highest correlation features, MFCC,
\WavAE, and \WavCCCAE for the evaluation of head-motion prediction.
Table~\ref{tab:rms_cca} shows the comparison of different systems,
where the quality of predicted head motion was evaluated in terms of
NMSE and local CCA with the ground truth (original head motion). 
We also computed local CCA between the ground truth and randomised sequences of another subject different from Subjects A and B to estimate a chance score for the two original head-motion sequences that are totally different and unsynchronised from each other and supposed to have no correlations. The estimated chance score for Subject A, Subject B and IEMOCAP respectively is $\rho_A = 0.16$, $\rho_B =0.11$ and $\rho_{IEMOCAP} =0.18$.

Although \ModelAE shows the lowest NMSE on the test set, it is just because the predicted head motion had little
movement,  which resulted in NMSE being close to 1.0. This is also reflected in the local CCA that,
\ModelAE is worse than the chance score for both subjects. 
\ModelCCCAE gets performance comparable to \ModelMFCC in terms of NMSE.
\ModelMFCC gets the highest local CCA.
Overall, the quality of \ModelMFCC and \ModelCCCAE in the test dataset is higher than the chance score.

\begin{figure}[t]
\begin{center}
\centerline{
\includegraphics[width=\linewidth, height=5cm, keepaspectratio]{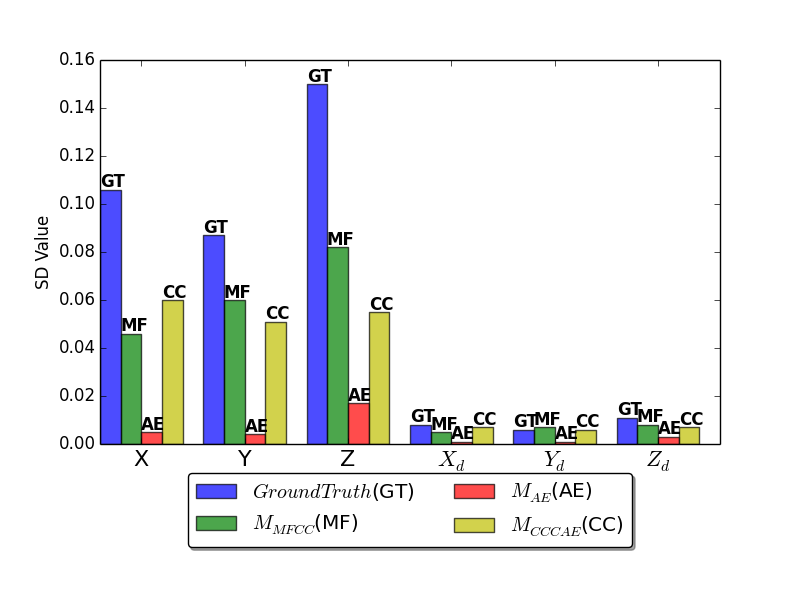}}
\vskip -3mm
\caption{
  Comparison of predicted head motion trajectories with different models in terms of
  standard deviation of each rotation parameter ($X,Y,Z$) and
  its derivative, where values are averaged over the two
  subjects.
}
\label{fig:std}
\end{center}
\vskip -10mm
\end{figure}

CCA captures only one aspect of similarity, i.e., linear correlations
between two data streams, and it does not tell us how similar the two
streams are in terms of other aspects such as dynamic range and
smoothness, which we believe are also crucial factors in human
perception. 
We thus calculated the standard deviation (SD) of
each head motion trajectory and its derivative, i.e., velocity, whose result is shown in 
Figure~\ref{fig:std}. \ModelAE has the smallest SD in all trajectories and velocities, which confirms that \ModelAE has very little movement as mentioned above. The ground truth has the largest SD  over trajectories. 
\ModelMFCC and \ModelCCCAE show comparable performance, but yet not close enough to the ground truth. Compared in terms of velocity,  the ground truth, \ModelMFCC, and \ModelCCCAE are similar to each other. This indicates that they are likely to have the same level of smoothness.

\subsection{Subjective Evaluation}
We conducted a perceptual test using the MUltiple Stimuli with Hidden Reference and Anchor (MUSHRA)\cite{mushra}. 
We developed five test groups from Subject A, where each test group consisted of 3 randomly selected audio samples in the test set, and animations were created from each sample using 5 models: Ground Truth, \ModelAE, \ModelCCCAE, \ModelMFCC, and Anchor. Each animation lasts $8-12$ seconds long (including speaking and listening frames). 
The anchor is created by selecting the original head motion of another speaker with different utterances. This ensures that the anchor head motion has a natural behaviour, but it does not synchronise with the audio. The evaluation is performed such that every participant is assigned one test group and the animations of each test group are shown in a random order. Then, each participant watches each head-motion animation and gives a score, between $0 - 100$, for each animation. A group of 20 participants were involved in this evaluation and they were asked to give a score to each animation according to the naturalness of the synthesised head motion.

\begin{figure}[t]
\begin{center}
\centerline{
\includegraphics[width=\linewidth, height=5cm, keepaspectratio]{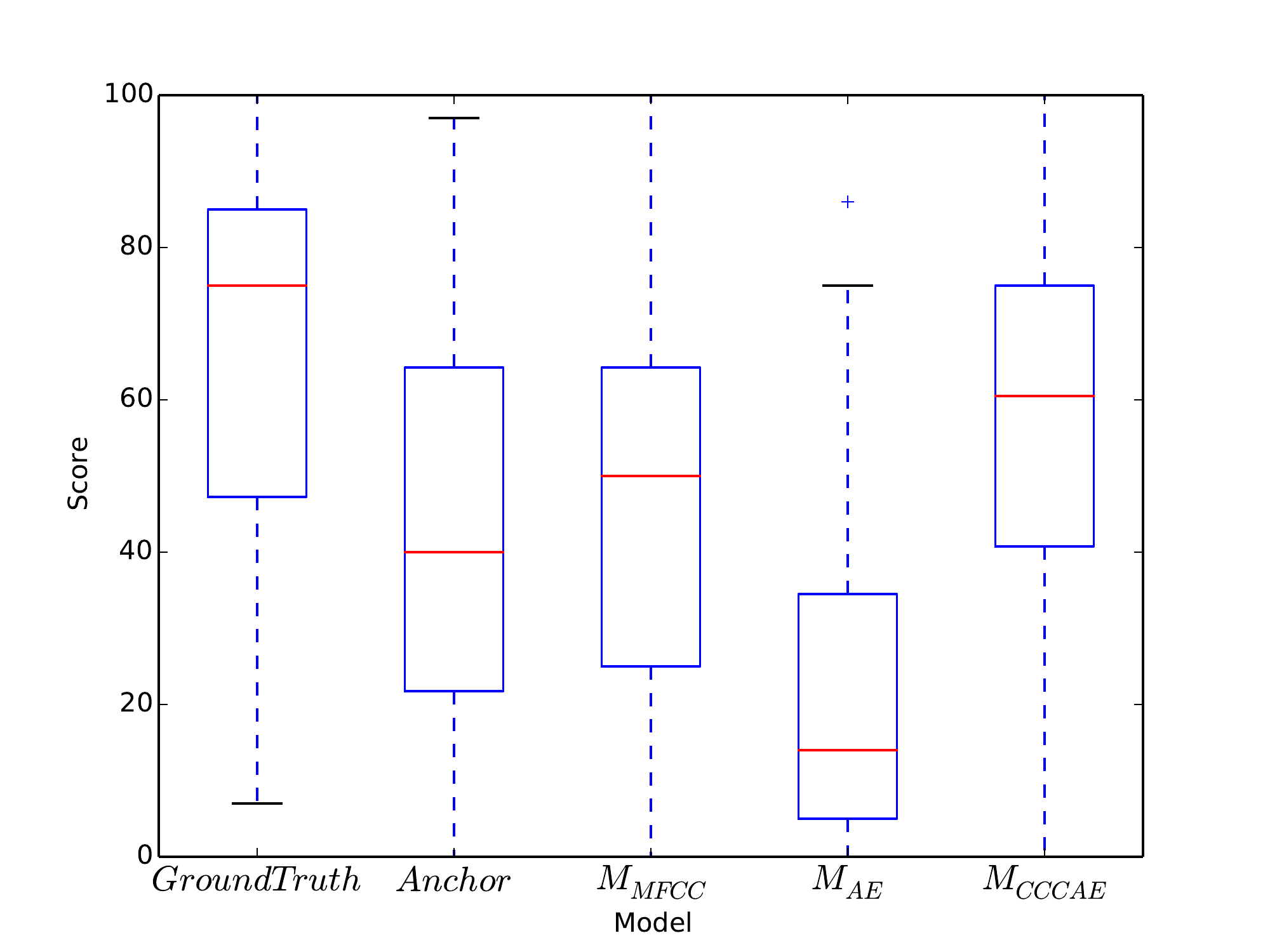}}
\caption{The Boxplot of the MUSHRA score for the Subject A's animation of each model - horizontal line indicates the median.}
\label{fig:subjective}
\end{center}
\vskip -10mm
\end{figure}

The result is shown in Figure~\ref{fig:subjective}. \ModelAE scored the lowest among all including the anchor. We think the reason could be that as the predicted head motion with \ModelAE conducted a relatively minor movement, which may seem contrary to regular human beings' behaviour, from the participants' perspective. The anchor scored the second lowest as expected, participants were able to figure out the non-synchronicity between the head motion and audio. Compared between \ModelMFCC and \ModelCCCAE models, participants scored higher for \ModelCCCAE.


One explanation for \ModelMFCC achieving better in the objective, but lower in the subjective than \ModelCCCAE may be that the participants were affected by the active head motion generated by \ModelMFCC while the agent is listening, and they felt against natural human being. This also shows that objective is in things quantifiable, whereas subjective is one open to greater interpretation based on personal feeling\cite{leahu2008}.

\section{Conclusions}
In this paper, we have proposed an approach to create a highly correlated feature with head motion from raw waveform data using CCCAE. From the objective evaluations, we can conclude that (1) CCCAE enables creation of a more correlated feature (\WavCCCAE) with the head motion than \WavAE and other popular spectral features such as MFCC and FBank. (2) the \ModelCCCAE achieved the lowest NMSE in test dataset, although the local CCA is not the highest. (3) the analysis based on SD shows that \ModelMFCC and \ModelCCCAE are comparable performance. (2) and (3) indicate that \WavCCCAE is capable of being used in achieving state-of-the-art results for predicting natural head motion with the advantage of the CCCAE. (4) MUSHRA test shows that excluding the ground truth, participants preferred to choose the animation generated by \ModelCCCAE over the others. Overall, our \ModelCCCAE shows better performance than \ModelMFCC. 

\bibliographystyle{IEEEtran}

\bibliography{mybib}


\end{document}